\documentclass{article}
\usepackage{arxiv} 

\makeatletter
\newdimen\xfigwd
\xfigwd=\textwidth
\makeatother
\usepackage{amsmath,amssymb,amsfonts,tabularx}
\usepackage[hidelinks]{hyperref}
\usepackage{cite}
\usepackage{url}
\usepackage{paralist}
\usepackage{flushend}
\usepackage{graphicx}
\usepackage{makecell}
\usepackage[table, dvipsnames]{xcolor}
\usepackage{array,multirow}
\usepackage{xspace}
\graphicspath{{figures/}}
\usepackage{adjustbox, colortbl}
\usepackage{algorithm}
\usepackage{algpseudocode}
\usepackage{booktabs}
\usepackage{placeins}
\usepackage{mdframed}

\newenvironment{remarkbox}%
{%
  \begin{mdframed}[%
    backgroundcolor=gray!10,
    linecolor=white,
    linewidth=0pt,
    skipabove=\topskip,
    skipbelow=\topskip
  ]
  \bfseries TAKEAWAY.~\normalfont
}%
{%
  \end{mdframed}
}

\newenvironment{takeawaybox}%
{%
  \begin{mdframed}[%
    backgroundcolor=blue!10,
    linecolor=white,
    linewidth=0pt,
    skipabove=\topskip,
    skipbelow=\topskip,
    innerleftmargin=6pt,
    innerrightmargin=6pt,
    innertopmargin=6pt,
    innerbottommargin=6pt
  ]
  \textbf{REMARK.}\ \normalfont
}%
{%
  \end{mdframed}
}

\title{Toward Realistic Adversarial Attacks in IDS: A Novel Feasibility Metric for Transferability}

\author{
  Sabrine Ennaji \\
  Sapienza University of Rome, Italy \\
  \texttt{ennaji@di.uniroma1.it}
  \And
  Elhadj Benkhelifa \\
  Staffordshire University, UK \\
  \texttt{e.benkhelifa@staffs.ac.uk}
  \And
  Luigi Vincenzo Mancini \\
  Sapienza University of Rome, Italy \\
  \texttt{mancini@di.uniroma1.it}
}

\begin{document}

\maketitle

\begin{abstract}
Transferability-based adversarial attacks exploit the ability of adversarial examples, crafted to deceive a specific source Intrusion Detection System (IDS) model, to also mislead a target IDS model without requiring access to the training data or any internal model parameters. These attacks exploit common vulnerabilities in machine learning models to bypass security measures and compromise systems. Although the transferability concept has been widely studied, its practical feasibility remains limited due to assumptions of high similarity between source and target models. This paper analyzes the core factors that contribute to transferability, including feature alignment, model architectural similarity, and overlap in the data distributions that each IDS examines. We propose a novel metric, the Transferability Feasibility Score (TFS), to assess the feasibility and reliability of such attacks based on these factors. Through experimental evidence, we demonstrate that TFS and actual attack success rates are highly correlated, addressing the gap between theoretical understanding and real-world impact. Our findings provide needed guidance for designing more realistic transferable adversarial attacks, developing robust defenses, and ultimately improving the security of machine learning-based IDS in critical systems.
\end{abstract}


\section{Introduction}\label{sec:intro}
The increasing adoption of digital technologies has intensified the need for robust network security measures. Intrusion Detection Systems (IDS) play a crucial role in protecting modern networks by continuously monitoring and analyzing network traffic for identifying suspicious activities and mitigating data breaches \cite{ozkan2021comprehensive}. However, traditional IDS, which rely on static, signature-based detection strategies, fail to address more sophisticated and dynamic cyberattacks \cite{huang2023artificial}. This challenge has urged researchers to explore innovative methods, leading to the integration of machine learning (ML) mechanisms \cite{chou2021survey, lansky2021deep}. By incorporating ML, IDS have become more adaptive and capable of identifying both known and unknown forms of attacks \cite{cantone2024machine}. However, their reliance on learned patterns make them vulnerable to adversarial attacks, where inputs are slightly manipulated in a way to trick the IDS model into misclassify malicious activities as benign \cite{shu2020generative}. 

In black-box settings \cite{zhu2022black}, where attackers lack access to the internal workings of the target IDS (e.g., its architecture, parameters, and training data), the transferability concept becomes fundamental for adversarial strategies \cite{roshan2024black}. Transferability refers to the ability of adversarial examples generated for one model (the source) to effectively deceive another one (the target). This approach allows attackers to rely on substitute models, which are trained to mimic the target system, as proxies to create adversarial examples. Attackers benefit from the reliance on transferability because it avoids the need to interact with the target system, which may be monitored or highly secured \cite{alhajjar2021adversarial}. 

However, a variety of assumptions must be maintained for transferability-based attacks to be successful \cite{demontis2019adversarial}. First, the source and target models should demonstrate considerable similarity in both architecture and behavior. Second, feature alignment is important, which means that the features used by the source model must closely match those used by the target model. Lastly, transferability ensures that carefully crafted adversarial inputs continue to work in both systems by assuming that the training data of the source and target models have overlapping data distributions.

In real-world settings, these assumptions are often unrealistic due to the significant variability in IDS deployments \cite{khraisat2021critical}. For example, IDS models might vary significantly in their architecture, feature extraction techniques, or even the types of attacks they have been designed to identify. Additionally, companies might train their IDS on proprietary data, which would produce different data distributions that differ substantially from publicly accessible datasets \cite{yusof2022visualizing}. 

Although such attacks may be challenging to execute in real-world settings, they remain a potential threat under specific conditions. Consequently, a systematic evaluation of the feasibility of transferability in realistic scenarios is essential; not only to assess the viability of such attacks but also to guide researchers in addressing this limitation.

\textbf{MOTIVATION.} 
In existing research, transferability is mainly evaluated in controlled settings, assuming optimal conditions such as feature matching and data similarity. They ignore the variability present in practical IDS environments, including diverse feature sets, architectures and network conditions. The disconnect between theoretical research and practical applications illustrates the necessity of a more thorough investigation into the feasibility and actual limitations of transferability. 

\textbf{CONTRIBUTIONS.} 
This paper addresses the limitations in current research by providing a comprehensive analysis of transferability-based attacks on ML-based IDS. Our contributions are summarized as follows: 

\begin{itemize} 
\item \textbf{Limitations analysis:} We thoroughly investigate the main challenges to transferability, such as feature misalignment, architectural disparities, and data distribution differences. 
\item \textbf{Transferability Feasibility Score (TFS):} We propose a novel metric to systematically assess the practicality of transferability-based attacks under real-world conditions, allowing researchers to quantify the feasibility of their methods. 
\item \textbf{Insights for defenses:} Based on our analysis, we provide actionable recommendations for IDS designers to improve feature variability, architectural diversity and other resilience measures to reduce transferability risks.
\end{itemize}

\textbf{PAPER STRUCTURE.} The remainder of this paper is organized as follows: Section \ref{sec:review} reviews the background and related work on transferability-based adversarial attacks and their assumptions. Section \ref{sec:methododlogy} present our methodology, including the development of the Transferanbility Feasability Score (TFS) and the experimental setup for assessing transferability limitations in realistic IDS settings. Section \ref{sec:discussion} discusses the achieved results, focusing on the impact of feature misalignment, architecture diversity, and data variability on transferability.
Finally, Section \ref{sec:conclusions} concludes the paper by summarizing our findings and insights.

\vspace{1cm}
\section{Background and Related Work}
\label{sec:review}
\subsection{Overview of Adversarial Attacks}
Adversarial attacks generate inputs aimed at deceiving the classification process in order to exploit vulnerabilities in intelligent models. Attackers attempt to avoid detection in the context of ML-based IDS by subtly modifying network traffic patterns through methods that are frequently undetectable.
They involve the generation of perturbations $\delta$ such that a crafted input $x' = x + \delta$ leads to misclassification by the IDS. \\
For a given model $f(x)$, an adversarial attack aims to increase the probability of misclassification:

\[
\underset{\delta}{\text{argmax}} \, \mathcal{L}(f(x'), y) \quad \text{subject to} \quad ||\delta||_p \leq \epsilon \; \; \; \; (1),
\]

where $\mathcal{L}$ is the loss function, $y$ is the true label, and $\epsilon$ constrains the perturbation under a specific norm $||\cdot||_p$ (e.g., $L_2$ or $L_\infty$). 

For ML-based IDS, $\delta$ might require modifying traffic features (e.g., packet size, flow duration, etc.) to bypass detection.

These attacks can be categorized based on the knowledge of the attacker: \textit{white-box} attacks, where the attacker has full knowledge of the model $f(x)$, and \textit{black-box} attacks, where $f(x)$ is unknown, necessitating indirect methods such as transferability or substitute modeling to detect vulnerabilities.

\subsection{Transferability in Adversarial Attacks}
Transferability was first explored by Goodfellow et al. \cite{papernot2016transferability} based on the concept that adversarial examples generated for a source model $f_s(x)$ can effectively mislead a target model $f_t(x)$ into making wrong predictions, even when the attacker has no direct access to $f_t$. This scenario is illustrated in Figure \ref{fig:trs} and can be described as follows:

\begin{align*}
\text{if} \, \mathcal{L}(f_s(x'), y) > \mathcal{L}(f_s(x), y) \ \; \;  (2), \\
\text{then} \, \mathcal{L}(f_t(x'), y) \approx \mathcal{L}(f_t(x), y) \; \;  (3),
\end{align*}

The adversarial example $x'$, generated to deceive the source model $f_s(x)$, is assumed to have a similar effect on the target model $f_t(x)$.

\subsubsection{Factors Influencing Transferability}

The effectiveness of transferability-based attacks relies on the following core factors:

\begin{figure*}[h!]
    \centering
    \includegraphics[width=\textwidth]{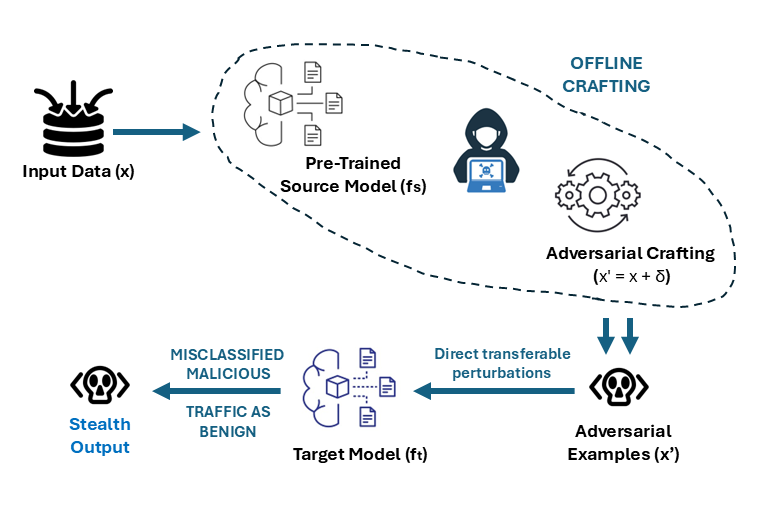}
    \caption{\parbox{0.7\textwidth}{\raggedright
    \textbf{Transferability-based Attacks against ML-based NIDS:} The attack starts with benign input (\(x\)) and uses a source model (\(f_s\)) (pretrained or public proxy) to craft adversarial examples (\(x' = x + \delta\)) offline. Perturbations (\(\delta\)) are designed to be adversarial while maintaining realism, then applied to the target model (\(f_t\)) without direct interaction, causing misclassification and successful evasion.
    }}
    \label{fig:trs}
\end{figure*}

\begin{enumerate}
\item \textbf{Feature alignment:} The probability of transferability is increased when the source model $f_s(x)$ and the target model $f_t(x)$ have overlapping feature spaces and comparable feature importance. Adversarial scenarios are less likely to succeed if $f_s(x)$ and $f_t(x)$ rely on different features or assign the same features different weights.

\item \textbf{Architecture similarity:} When the source and target models have comparable architectures, hyperparameters, or learning processes, transferability performs optimally. For instance, models with essentially different architectures (such as a neural network Vs. a decision tree) are less likely to demonstrate transferability than two convolutional neural networks trained on the same dataset.

\item \textbf{Data distribution:} Transferability assumes that adversarial examples $x'$ lie within the high-density regions of the data distributions of the source and target models. If the source and target models are trained on different datasets or experience significant domain shifts ($P_s(x) \neq P_t(x)$), the effectiveness of transferability decreases.
\end{enumerate}

Although transferability-based attacks are theoretically appealing, they are less effective in heterogeneous environments where conditions such as aligned feature spaces ($F_s \neq F_t$) or similar data distributions ($P_s(x) \neq P_t(x)$) are not met, reducing their success rates.

\subsubsection{Attack detectability}

The evasiveness of transferability-based attacks lies on their reliance on external source models instead of direct interactions with the target system. This approach minimizes the  visibility of the attacker's behaviour, making it less detectable than other black-box methods, which are categorized into four approaches as shown in Figure \ref{fig:bb}. The key factors contributing to this stealthiness are:

\begin{figure*}[h!]
    \centering
    \includegraphics[trim=0cm 6cm 0cm 0cm, clip, width=1\textwidth]{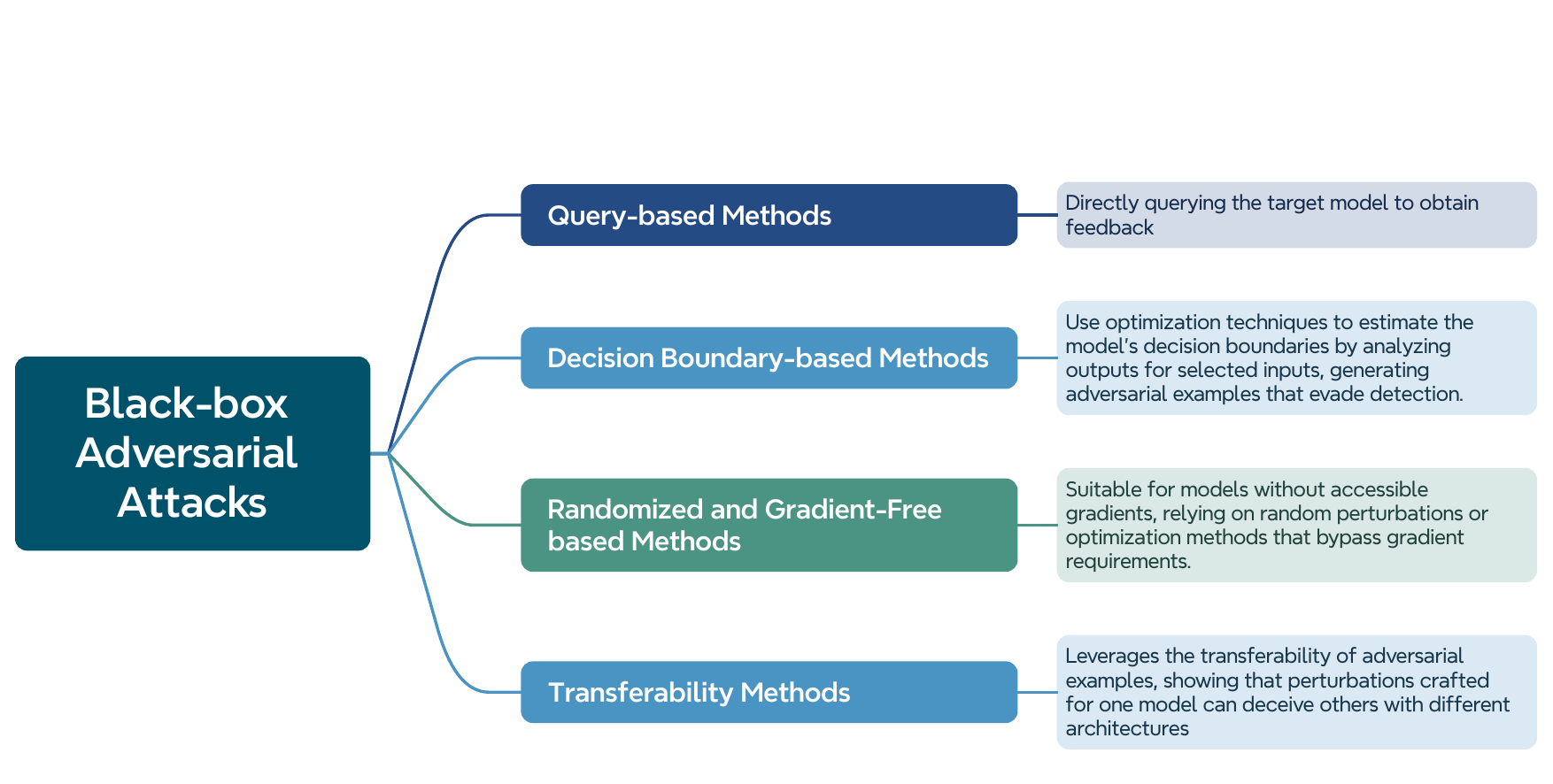}
    \caption{Categories of Black-box Adversarial Attacks}
    \label{fig:bb}
\end{figure*}

\begin{enumerate}
\item \textbf{No query footprint:} Transferability-based attacks do not necessitate querying the target model to generate adversarial perturbations. On the other hand, query-based attacks rely on repeatedly sending crafted inputs to the target system and examining its outputs (e.g., labels, confidence scores, or probabilities). This iterative querying process often leaves detectable traces, such as:
\begin{itemize}
\item \textit{Anomalous interaction patterns:} Logs could indicate a high frequency of related or well-crafted queries, particularly if the target system monitors traffic for suspicious activity.
\item \textit{Rate-limiting alerts:} Excessive interaction attempts are easily noticeable, as many systems employ throttle mechanisms or query restrictions.

\item \textit{Signature-based detection:} Query patterns associated with known attack types can trigger alarms in sophisticated IDSs.
\end{itemize}

Transferability-based attacks considerably limit the possibility of being detected or prevented during execution by eliminating the requirement for queries and avoiding the creation of such detectable traces.

\item \textbf{Pre-Existing models and data:}
Adversarial examples created with pre-existing models and datasets, which serve as substitutes for the target system, are essential to transferability. By using these external resources, attackers can create adversarial inputs without interacting with the target system, avoiding detection mechanisms that might otherwise identify suspicious activity. This strategy provides a variety of advantages:

\begin{itemize}
\item \textit{Source models as substitutes:} Models that approximate the target system can be used by attackers and are either independently trained or publically available. An efficient alternative to producing adversarial examples is a model that has been trained on a similar dataset or domain.

\item \textit{Public datasets for training:} Attackers often exploit publicly available datasets to train and evaluate their substitute models \cite{de2023survey}. This reduces the requirements for direct access to the target system's private data, further reducing the attack’s visibility.

\item \textit{Offline adversarial crafting:} Every adversarial example is created and tested offline, away from the target setting. This makes it significantly more difficult for defenders to identify or trace the attack's source because it ensures that no information from the target system is required until the attack is launched.

\end{itemize}

\end{enumerate}

\begin{remarkbox}
Transferability context remains a critical area of focus because it avoids direct queries to the target model, making it less detectable and more appropriate to black-box scenarios.
\end{remarkbox}

\subsection{Limitations of Existing Studies}
Adversarial attacks on ML-based IDS have gained excessive attention due to the growing reliance on intelligent models in sensitive applications \cite{rajkumar2025multi, lin2025enhancing} (e.g., smart cities, healthcare systems, banking, etc.\cite{vinayakumar2019deep, hady2020intrusion}). The development and evaluation of attack methods, in particular the transferability of adversarial examples in black-box settings, have been the subject of extensive research. Although these studies reveal the theoretical weakness of these systems, transferability-based attacks are still not highly feasible in a variety of heterogeneous IDS setups. This limitation prevents  a deeper understanding of this evolving field and restricts the development of strong adversarial defenses able to generalize effectively across different real-world scenarios.
For instance, Debicha et al. \cite{debicha2023adv} propose a black-box adversarial attack framework based on transferability strategy to bypass ML-based IDS. While the findings successfully show that adversarial samples can minimize detection rates, the feasability of the proposed attack is limited in real-world deployments. The study ignores the challenges that arise from noisy traffic data, dynamic and proprietary feature sets, and sophisticated IDS defenses (e.g., adversarial training, ensemble learning etc.). Despite the high attack success rate achieved in the controlled settings, the findings lack generalizability to practical scenarios, where transferability is significantly reduced by network heterogeneity and adaptive defenses. Therefore, this approach might not be able to address the complexities of real-world IDS deployments. 

Similarly, Zhang et al. \cite{zhang2024explainable} propose the Explainable Transfer-based Adversarial Attack (ETA) architecture using an ensemble substitute model that includes both differentiable and non-differentiable components. It applies a min-max approach to optimize transferability and their Importance-Sensitive Feature Selection (ISFS) method, which combines perturbation interpretations and cooperative game theory, to select non-robust features. While this approach shows promising attack success rates across different datasets, it assumes the target and substitute models have sufficiently similar data distributions, feature importance rankings, and decision boundaries. However, it is challenging to verify these assumptions in practical settings. For example, the ISFS method selects non-robust features based on cooperative game theory, assuming alignment between substitute and target feature importance, which fails under partial feature knowledge. Moreover, using substitute datasets that approximate the target distribution presumes access to aligned data, which is rarely feasible in real-world scenarios. 

Moreover, it has been proposed, in \cite{chale2023constrained}, to use a surrogate model, specifically a 1-dimensional CNN, to generate adversarial examples in the limited cyber domain. This strategy uses a meta-heuristic optimization strategy to create adversarial examples. It perturbs raw packet payloads, preserving their functionality, while maximizing cross-entropy loss with respect to a surrogate model. Then,  these examples are transferred to three different target NIDS models (CNN, FNN, and Adaboost) to evaluate their evasion rates. The authors assume shared architecture, feature alignment, and data distribution between the surrogate and target models, which enhance transferability but limit the applicability of their method in strict black-box settings. 

In another recent study \cite{zhang2024toward}, the authors explore transferability of adversarial examples against autoencoder-based network IDS. They adapts gradient-based and optimization-based attacks (e.g., FGSM, BIM, PGD, etc.), while preserving network protocol integrity. Their main contribution lies in the use of a Linear Autoencoder (LAE) as a substitute model, which simplifies the autoencoder by removing activation functions, thereby improving the transferability of adversarial examples to other autoencoder-based models (e.g., DAGMM, SAE, and VAE). By focusing on the linearity of the surrogate model, this approach leads to a significant improvement in black-box attack success rates. However, its effectiveness and applicability in strict black-box conditions is limited by its reliance on partial knowledge of normal samples and feature extractors. Furthermore, the approach's assumptions of architectural similarity, feature alignment, and shared data distribution between target and substitute models, may not always hold in realistic settings.

In addition, Roshan et al. \cite{roshan2024black} examine black-box adversarial transferability within a cyber attack detection system using deep learning models. They trained a surrogate and a target model on the same CICDDoS-2019 dataset and generated adversarial perturbations using the Fast Gradient Sign Method (FGSM). The aim was to evaluate how well adversarial examples crafted on the surrogate model could trick the target model without direct access to its internal architecture. Although their approach shows successful transferability, the assumption that the surrogate and target models are trained on the same datasets and the reliance on gradient-based perturbations represent a controlled environment rather than a realistic black-box situation. Such aligned data distributions or model structures are rarely available to attackers in practice, which severely restricts the applicability of the suggested approach in real-world scenarios.

Alhussien et al. \cite{alhussien2024constraining} explored the impact of adversarial attacks on ML and DL-based NIDS by proposing a novel set of domain-specific constraints to create adversarial examples. Their approach focused on maintaining the validity of the perturbed adversarial traffic while preserving the statistical and semantic relationships between traffic features. Although the study showed improved attack success rates under constrained conditions, the effectiveness of the attack was highly dependent on the similarity between the surrogate and target models in terms of feature space and data distribution. This reliance on similarity raises concerns about the approach's applicability in real-world black-box settings, where attackers usually do not have access to the target model's internal structure and training data. Additionally, the study ignores real-world complexities including evolving model architectures, dynamic network traffic patterns, and the existence of defense systems, which may limit the generalizability of the findings to practical scenarios.

Recently, Adeke \cite{adeke2025investigating} developed a surrogate-target model framework with the zeroth-order optimization (ZOO) technique to craft adversarial examples, relying on the target model’s observable behavior. Despite the consideration of the black-box nature of the attack, feature-based analysis, and evaluation on practical datasets (IoT-23 and UNSW-NB15), the authors assume identical training data for surrogate and target models. Given that real-world data distributions frequently vary, this approach seems unrealistic. Furthermore, the controlled setting ignores adaptive defenses, real-time data variability, and evolving network conditions, which limits the generalizability of the obtained results.
 
\begin{remarkbox}Existing transferability attacks are limited by their reliance on surrogate-target model similarities. Further studies should focus on reducing these dependencies, to improve their real-world impact.
\end{remarkbox}

\section{Proof of Concept}
\label{sec:methododlogy}
To address the limitations of prior research \cite{debicha2023adv, zhang2024explainable, chale2023constrained, zhang2024toward}, which frequently ignore the interplay between feature alignment, architectural similarity, and data distribution homogeneity, this section describes the methodology used to assess the feasibility of transferability-based adversarial attacks in black-box scenarios. Specifically, we reproduce the methodology from a previous study \cite{debicha2023adv} and evaluate its transferability using our suggested Transferability Feasibility Score (TFS). Given the studies in the literature are built on the same assumption, reproducing one representative approach is sufficient to evaluate broader limitations. This allowed us to focus on examining the feasibility of transferability in real-world situations while providing insights into its practical restrictions using TFS.
 
\subsection{Transferability Feasibility Score (TFS)}

The main contribution of this study is the proposition of a novel metric, namely; Transferability Feasibility Score (TFS). It provides a quantitative measure to evaluate the feasibility of transferability-based adversarial attacks. It incorporates three core dimensions: feature alignment (\(f_{\text{align}}\)), architectural similarity (\(A_{\text{sim}}\)), and data distribution homogeneity (\(D_{\text{hom}}\)); which collectively define the probability of successful transferability between surrogate and target models.  Algorithm 1 provides a concise overview of the proposed TFS for quick reference.

 Formally, TFS is defined as follows:
\[
TFS = \alpha \cdot f_{\text{align}} + \beta \cdot A_{\text{sim}} + \gamma \cdot D_{\text{hom}} \;\;\; (4)
\]
where
\begin{enumerate}

\item (\(f_{\text{align}}\)) measures the overlap and compatibility of the feature spaces between the surrogate and target models.
\[
f_{\text{align}} = \frac{|F_{\text{s}} \cap F_{\text{t}}|}{|F_{\text{s}} \cup F_{\text{t}}|} \;\;\; (5),
\]
\(F_{\text{s}}\) and \(F_{\text{t}}\) are the feature sets used by the surrogate and target models, respectively. It is the \textbf{Jaccard Similarity Inde}x \cite{hwang2018new}, a commonly used statistic to assess how similar two sets are.

\begin{itemize}
\item  \( |F_s \cap F_t| \) counts the features that both models have in common.
\item \( |F_s \cup F_t| \) represents all of the distinct features identified in both models.
\end{itemize}

\begin{remarkbox}
A higher \(f_{\text{align}}\) scores indicates greater feature similarity but lower real-world feasibility.
\end{remarkbox}

\item (\(A_{\text{sim}}\)) measures how similar the structures of the surrogate and target models are.
\[
A_{\text{sim}} = 1 - \frac{\|P_{\text{s}} - P_{\text{t}}\|_2}{\|P_{\text{t}}\|_2},
\]
\(P_{\text{s}}\) and \(P_{\text{t}}\) are the hyperparameter vectors of the surrogate and target models.

\begin{itemize}
\item $\|P_s - P_t\|_2$ measures the \textbf{Euclidean distance} \cite{elmore2001euclidean} \textit{(L2 norm)} between the target (\(P_t\)) and surrogate (\(P_s\)). It quantifies the model distance in parameter space.  
\item $\|P_t\|_2$ normalizes the distance relative to the target model's parameters magnitude. This guarantees that the similarity is scaled appropriately, because models with superior parameter magnitudes may otherwise bias the measure.
\item The \(1 -\) ensures the value has been altered to show similarity instead of distinctions.
\end{itemize}

\begin{remarkbox}\(A_{\text{sim}}\) ranges from 0 to 1, where 1 indicates identical architectures.
\end{remarkbox}
\item (\(D_{\text{hom}}\)) measures the similarity of the datasets used for training the surrogate and target models.
\[
D_{\text{hom}} = \frac{1}{1 + \text{Wasserstein}(D_{\text{s}}, D_{\text{t}})} \;\;\; (6),
\]
\(\text{Wasserstein}(D_{\text{s}}, D_{\text{t}})\) is \textbf{Wasserstein distance} \cite{panaretos2019statistical} between the distributions \(D_{\text{s}}\) (surrogate) and \(D_{\text{t}}\) (target).

\begin{itemize}
\item Wasserstein($D_s, D_t$) measures the transformation cost between distributions.
\item Adding $1$ in the denominator prevents undefined or negative metric values for identical distributions.
\item The reciprocal $\frac{1}{1 + \text{Wasserstein}(D_{\text{s}}, D_{\text{t}})}$ inverts the distance to quantify homogeneity.
\end{itemize}

\begin{remarkbox} Smaller distances indicate greater homogeneity.
\end{remarkbox}

\item  \(\alpha\), \(\beta\), and \(\gamma\) are weighting coefficients, which represent the relative significance of each dimension. To determine the optimal coefficients, the process requires minimizing the sum of squared errors (SSE) between the observed attack success rate (\( y \)) and the predicted rate based on the factors (\( f_{\text{align}}, A_{\text{sim}}, D_{\text{hom}} \)), as demonstrated in the following formula.

\[
\text{SSE} = \sum_{i=1}^n \left( y_i - (\alpha \cdot f_{\text{align}_i} + \beta \cdot A_{\text{sim}_i} + \gamma \cdot D_{\text{hom}_i}) \right)^2  (7),
\]

To achieve this, we employ the normal equation method for linear regression:

\[
\mathbf{X}^\top \mathbf{X} \mathbf{w} = \mathbf{X}^\top \mathbf{y} \;\;\; (8),
\]

Where:

\[
\mathbf{X} = 
\begin{bmatrix}
f_{\text{align}_1} & A_{\text{sim}_1} & D_{\text{hom}_1} \\
f_{\text{align}_2} & A_{\text{sim}_2} & D_{\text{hom}_2} \\
\vdots & \vdots & \vdots \\
f_{\text{align}_n} & A_{\text{sim}_n} & D_{\text{hom}_n}
\end{bmatrix}
\]

\[
\mathbf{w} = 
\begin{bmatrix}
\alpha \\
\beta \\
\gamma
\end{bmatrix}, \quad
\mathbf{y} = 
\begin{bmatrix}
y_1 \\
y_2 \\
\vdots \\
y_n
\end{bmatrix}
\]

\begin{itemize}
    \item \( \mathbf{X} \) is the matrix of independent variables (\( f_{\text{align}}, A_{\text{sim}}, D_{\text{hom}} \)).
    \item \( \mathbf{w} \) presents the coefficients vector (\( \alpha, \beta, \gamma \)).
    \item \( \mathbf{y} \) is the attack success rate vector.
\end{itemize}

The analytical solution for the coefficients is solved by:

\[
\mathbf{w} = (\mathbf{X}^\top \mathbf{X})^{-1} \mathbf{X}^\top \mathbf{y} \;\;\; (9),
\]

This procedure ensures that the coefficients optimize the linear model by considering the contribution of each factor to the observed attack success rate.

\end{enumerate}

\begin{figure*}[t!]
    \centering
    \includegraphics[width=\textwidth]{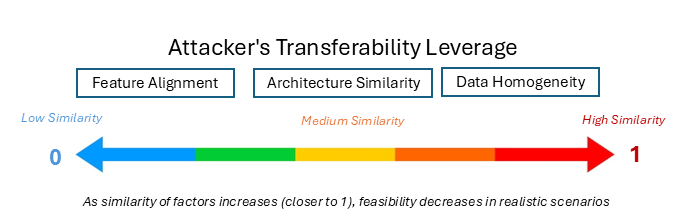}
    \caption{\parbox{0.85\textwidth}{\raggedright
    \textbf{Impact of Increasing Similarity on Transferability Feasibility in Realistic Scenarios}---Attacker's Transferability Leverage:
    As similarity of factors (Feature Alignment, Architecture Similarity, and Data Homogeneity) increases from low (0) to high (1), transferability feasibility decreases in realistic scenarios.}}
    \label{fig:feasibility}
\end{figure*}

\begin{remarkbox}
\begin{itemize}

\item \textbf{High TFS} (\( TFS \to 1 \)): Limited real-world applicability due to unrealistic assumptions, implying \(f_{\text{align}} \to 1\), \(A_{\text{sim}} \to 1\), and \(D_{\text{hom}} \to 1\). 

\item \textbf{Low TFS} (\( TFS \to 0 \)): Effective approach under practical conditions, resulting from low values in \(f_{\text{align}}\), \(A_{\text{sim}}\), or \(D_{\text{hom}}\). 
\end{itemize}
Figure \ref{fig:feasibility} depicts how the transferability feasibility decreases in realistic scenarios as the similarity of factors (Feature Alignment, Architecture Similarity, and Data Homogeneity) increases from low (0) to high (1).
\end{remarkbox}

\begin{algorithm}[t]
\caption{Transferability Feasibility Score (TFS) with Regression-Based Coefficients}
\label{alg:TFS}
\begin{algorithmic}[1]

\Require 
\begin{itemize}
  \item $F_s, F_t$: Feature sets (source, target)
  \item $P_s, P_t$: Architecture parameters (source, target)
  \item $D_s, D_t$: Data distributions (source, target)
  \item $y$: Observed (empirical) attack success rates
  \item Constraint: $\alpha + \beta + \gamma = 1$
\end{itemize}
\Ensure $TFS$: Transferability Feasibility Score

\Statex

\State \textbf{Step 1: Compute Factor Values}
\State $f_{\text{align}} \gets \dfrac{|F_s \cap F_t|}{|F_s \cup F_t|}$
\State $A_{\text{sim}} \gets 1 \;-\; \dfrac{\|P_s - P_t\|_{2}}{\|P_t\|_{2}}$
\State $D_{\text{hom}} \gets \dfrac{1}{1 + \mathrm{Wasserstein}(D_s, D_t)}$

\Statex

\State \textbf{Step 2: Formulate Regression Problem}
\State Construct $X \in \mathbb{R}^{n \times 3}$ where each row contains 
       $\bigl(f_{\text{align},i},\, A_{\text{sim},i},\, D_{\text{hom},i}\bigr)$ 
\State Form vector $y \in \mathbb{R}^n$ with the observed success rates $y_i$

\Statex

\State \textbf{Step 3: Solve for Regression Coefficients}
\State Compute $w = [\,\alpha,\; \beta,\; \gamma\,]^\mathsf{T}$ using:
  \[
    w \;=\; (X^\mathsf{T}X)^{-1} X^\mathsf{T}y
  \]

\Statex

\State \textbf{Step 4: Compute TFS}
\State $TFS \gets \alpha \cdot f_{\text{align}} \;+\; \beta \cdot A_{\text{sim}} \;+\; \gamma \cdot D_{\text{hom}}$

\Statex

\State \textbf{Return} $TFS$

\end{algorithmic}
\end{algorithm}

By providing a comprehensive assessment of transferability, our proposed Transferability Feasibility Score (TFS) evaluates the practicality and effectiveness of adversarial attacks in real-world scenarios and helps researchers identify and refine their approaches to the development of more robust and adaptive defenses.

\begin{remarkbox}
To the best of our knowledge, no existing feasibility metrics for adversarial transferability have been proposed. Performance-based indicators like Attack Success Rate (ASR) or Model Confidence Degradation are commonly used to evaluate current transferability-based attacks. Although these metrics measure the effectiveness of an attack, they do not evaluate whether the attack was conducted in real black-box conditions or if it relies on unrealistic surrogate-target similarities.
\end{remarkbox}

\subsection{Experimental Setup}
\subsubsection{CSE-CIC-IDS2018 Dataset}
To evaluate our proposed TFS metric and highlight limitations in existing transferability-based adversarial attack research, we adopt the experimental setup from \cite{debicha2023adv}, which employed different datasets (CTU-13 and CSE-CIC-IDS2018). To simplify our analysis, we restrict our experiments to the CSE-CIC-IDS2018. 
It is considered a sophisticated benchmark for evaluating IDS and was developed by the Canadian Institute for Cybersecurity (CIC) in collaboration with the Communications Security Establishment (CSE). 
CSE-CIC-IDS2018 is designed to mimic realistic network traffic and incorporates a wide range of both benign and malicious behaviors, which is suitable and sufficient to assess the practical feasibility of transferability-based adversarial attacks. Introducing additional datasets would not significantly change our findings, as current black-box attacks consistently rely on surrogate-target similarity across different scenarios.
Table \ref{tab:cse_cic_ids2018} presents a comprehensive summary of the CSE-CIC-IDS2018 dataset.

\subsubsection{Preprocessing}
In order to closely adhere to the paper's methodology, we preprocess the dataset as follows.
\begin{enumerate}
\item \textbf{Traffic filtering:} Since TCP traffic constitutes the majority of network traffic and is necessary for realistic attack scenarios, only this type of traffic is retained.
\item \textbf{Feature selection:} To focus only on network flow attributes, irrelevant columns have been removed such as \textit{Flow ID}, \textit{Src IP}, \textit{Dst IP},\textit{ Timestamp}, and other non-contributory metadata.
\item \textbf{Data cleaning:} Rows with missing values were removed to ensure data integrity.
\item \textbf{Feature engineering:} To improve the dataset, more features are extracted.
\begin{itemize}
    \item \textit{Bytes per second:} calculated by dividing the total number of bytes (forward and backward) by the flow duration.
    \item \textit{Packets per second:} calculated by dividing the total number of packets by the flow duration. To prevent division errors, zero-duration values are substituted with a minimal constant (1e-6).
\end{itemize}
\item \textbf{Normalization:} Min-Max scaling is employed to normalize all features in order to provide consistent ranges and machine learning algorithm compatibility.
\item \textbf{Data splitting:} The dataset was split into training (75\%) and testing (25\%) sets using stratified sampling to maintain the class distribution.
\end{enumerate}

\begin{table}[ht]
\centering
\caption{Summary of the CSE-CIC-IDS2018 Dataset}
\label{tab:cse_cic_ids2018}
\begin{tabular}{p{3cm}|p{12cm}}
\hline
\textbf{Attribute}           & \textbf{Description} \\ \hline
\textbf{Traffic Type}        & Realistic network traffic, including HTTP, HTTPS, FTP, SSH, and email communications \\ \hline
\textbf{Attacks}    & Brute Force (SSH, FTP), DoS (Slowloris, SlowHTTPTest, Hulk), DDoS (Botnet-based), Web Attacks (SQL injection, XSS), Infiltration (Unauthorized access), Botnet Traffic (Spam, Reconnaissance) \\ \hline
\textbf{Feature Categories}  & Basic Features (e.g., flow duration, total packets), Content Features (e.g., flags, flow stats), Time Features (e.g., inter-arrival times), Derived Features (e.g., bytes-per-second) \\ \hline
\textbf{Labeling}            & Instances labeled as benign or malicious, with specific attack types identified \\ \hline
\textbf{Volume of Data}      & Over 16 million records collected across 5 days \\ \hline
\textbf{ Total Features}  & 84 \\ \hline
\textbf{Pros}           & Diverse attack types, realistic traffic patterns, rich feature set \\ \hline
\textbf{Cons}         & May not represent all real-world attack types; synthetic environment introduces biases \\ \hline
\end{tabular}
\end{table}

\subsubsection{Transferability-based Attack Scenario}
Following the methodology of the attack scenario presented in the original study \cite{debicha2023adv}, we implemented the following procedure, formalized in Algorithm 2: 
\begin{itemize}
\item \textbf{Target model:} One of the target models used in the paper is a Random Forest (RF) classifier, which was set up with 200 estimators. It is trained using a carefully selected set of features that are modifiable and valuable in detecting network intrusions (i.e.. Duration, TotPkts, InBytes, OutBytes, BytesPerSec, PktsPerSec).

\item \textbf{Surrogate model:} To facilitate the generation of adversarial samples, the surrogate model is created to mimic the decision boundaries of the target RF-based IDS. Its architecture has been built as follows.
\begin{itemize} 
\item \textbf{Input layer:} It accepts a 4-dimensional input vector that matches the features that have been selected (Duration, TotPkts, InBytes, OutBytes).
\item \textbf{Hidden layer:} It includes three fully connected layers, each composed of 128 neurons. For non-linear transformations, a ReLU activation function is employed. Moreover, to mitigate the risk of overfitting, dropout layers are strategically placed after each hidden layer, with a dropout rate of 20\%.
\item \textbf{Output layer:} Dense layer comprising 2 neurons, using a softmax activation function. This configuration enables binary classification, predicting the probability of each class (i.e., Benign/Malicious).
\item \textbf{Optimization and loss function:} The model is optimized using the Adam optimizer with a learning rate of 0.001. The used loss function is Sparse Categorical Crossentropy, which is specifically created for integer-encoded labels, making it appropriate for this classification task.

\end{itemize}
\item \textbf{Feature selection and perturbation:} The adversarial scenario strategically targets a subset of features (i.e.,  duration, total packets, incoming bytes, and outgoing bytes) that are carefully selected to ensure that the perturbed network traffic maintains syntactically and semantically valid. The features selected for both the surrogate model and adversarial attack generation, with their descriptions, roles in the adversarial process, and applied constraints, are depicted in Table~\ref{tab:features_summary}.

\begin{algorithm}
\caption{Crafting Adversarial Examples for Flow-Based IDS}
\begin{algorithmic}[1]
\Procedure{CraftAdvEx}{$x$} \Comment{$x$ is a malicious  flow}
    \State $x_{\text{adv}} \gets x$ \Comment{Initialize adversarial example}
    \State $t \gets 1$ \Comment{Start iteration counter}
    \State $m \gets \text{mean\_difference()}$ \Comment{Alternative: mean\_ratio()}
    \Repeat
        \For{each mask $m_i \in \{\text{mask}_1, \ldots, \text{mask}_{15}\}$}
            \State $\epsilon \gets \text{sign}[\text{benign\_mean}(f) - x_0(f)] \cdot (c \cdot t) \cdot m(f)$
            \State $\epsilon \gets \epsilon \cdot m_i$ \Comment{Apply the mask}
            \State $x_{\text{adv}} \gets x_{\text{adv}} + \epsilon$ \Comment{Update adversarial flow}
            \State $x_{\text{adv}} \gets \text{Proj}(x_{\text{adv}})$ \Comment{Apply domain constraints}
            \If{$\text{predict}(x_{\text{adv}}) == \text{benign}$}
                \State \Return $x_{\text{adv}}$ \Comment{Adversarial example successfully crafted}
            \EndIf
        \EndFor
        \State $t \gets t + 1$ \Comment{Increment iteration counter}
    \Until{$\text{predict}(x_{\text{adv}}) == \text{benign}$}
\EndProcedure

\Procedure{Proj}{$x_{\text{adv}}$} \Comment{Apply domain constraints to ensure validity}
    \State $x_{\text{adv}} \gets \text{ApplySyntacticConstraints}(x_{\text{adv}})$
    \State $x_{\text{adv}} \gets \text{ApplySemanticConstraints}(x_{\text{adv}})$
    \State \Return $x_{\text{adv}}$
\EndProcedure
\end{algorithmic}
\end{algorithm}

\begin{table}[ht!]
\centering
\caption{Detailed Information on Perturbed Features}
\label{tab:features_summary}
\begin{tabular}{l|p{5cm}|p{4.5cm}|p{4cm}}
\hline
\textbf{Features} & \textbf{Description} & \textbf{Role in Adversarial Attack} & \textbf{Constraints Applied} \\ \hline
\textbf{Duration} &
Time duration of the network connection (in seconds). &
Perturbed to simulate longer or shorter connection durations. &
Must remain positive. Adjust \textit{BytesPerSec} and \textit{PktsPerSec} proportionally to maintain logical consistency. \\ \hline
\textbf{TotPkts} &
Total number of packets sent and received during the connection. &
Perturbed to simulate varying packet rates in the network traffic. &
Must remain positive. Adjust \textit{PktsPerSec} proportionally. \\ \hline
\textbf{InBytes} &
Total bytes received during the connection. &
Perturbed to simulate increased or decreased data inflow. &
Ensure \texttt{InBytes} is non-negative. Maintain consistency with \textit{OutBytes} for valid \textit{RatioOutIn}. \\ \hline
\textbf{OutBytes} &
Total bytes sent during the connection. &
Perturbed to simulate increased or decreased data outflow. &
Ensure \textit{OutBytes} is non-negative. Maintain consistency with \textit{InBytes} for valid \textit{RatioOutIn}. \\ \hline
\textbf{BytesPerSec} &
Derived feature: Total bytes per second during the connection (\( \frac{\text{TotLen Fwd Pkts} + \text{TotLen Bwd Pkts}}{\text{Duration}} \)). &
Adjusted based on perturbations to \textit{InBytes}, \textit{OutBytes}, and \textit{Duration} to maintain logical consistency. &
Ensure values remain finite and non-negative. \\ \hline
\textbf{PktsPerSec} &
Derived feature: Total packets per second during the connection (\( \frac{\text{TotPkts}}{\text{Duration}} \)). &
Adjusted based on perturbations to \textit{TotPkts} and \textit{Duration}. &
Ensure values remain finite and non-negative. \\ \hline
\textbf{RatioOutIn} &
Derived feature: Ratio of outgoing bytes to incoming bytes (\( \frac{\text{OutBytes}}{\text{InBytes} + \epsilon} \)). &
Perturbed indirectly by adjusting \textit{InBytes} and \textit{OutBytes} to simulate realistic data flows. &
Prevent division by zero (\( \epsilon > 0 \)). Ensure consistency between \textit{InBytes} and \textit{OutBytes}. \\ \hline
\end{tabular}
\end{table} 
Two main formulas control how features are perturbed.\\

 \textbf{\textit{Using mean ratios:}}
    \[
    \begin{aligned}
    x^{t}_{\text{adv}}(f) = \text{Proj}\bigg[ & x^{t-1}(f) + \text{sign}\big(\text{benign\_mean}(f) - x^0(f)\big) \\
    & \cdot (c \cdot t) \cdot \text{mean\_ratio}(f) \bigg] \;\;\;\; (10),
    \end{aligned}, 
    \]
    
 \textbf{\textit{Using mean differences:}}
    \[
    \begin{aligned}
    x^{t}_{\text{adv}}(f) = \text{Proj}\bigg[ & x^{t-1}(f) + \text{sign}\big(\text{benign\_mean}(f) - x^0(f)\big) \\
    & \cdot (c \cdot t) \cdot \left|\text{mean\_diff}(f)\right| \bigg] \;\;\;\; (11),
    \end{aligned}
    \]

Where:\\
 \(x^0(f)\): Original feature value.
 
 \(t\): Iteration number.
 
  \(c\): Scaling constant.
  
 \(\text{Proj}[\cdot]\): Constraint enforcement function.

\item \textbf{Projection and constraints:} To ensure the validity and realism of adversarial instances, the projection process involves both syntactic and semantic constraints.
\begin{itemize}
    \item \textbf{Syntactic constraints:} They maintain each feature's value within reasonable limitations; for example, they ensure sure that features like Duration and TotPkts are always positive and stay within realistic bounds. 
    
    \item \textbf{Semantic constraints:} They maintain the logical connections between interdependent features. For instance, dependent features like BytesPerSec and PktsPerSec are updated correspondingly to changes in Duration and TotPkts, whereas the RatioOutIn feature must show constant proportions between OutBytes and InBytes.
    
\end{itemize}

Adversarial samples are gradually adjusted through an iterative process, based on the formulas above with constraints applied after each step, ensuring the generated network remains valid and realistic. 

\end{itemize}

\begin{figure*} [h!]
    \centering
    \includegraphics[width=0.89\textwidth]{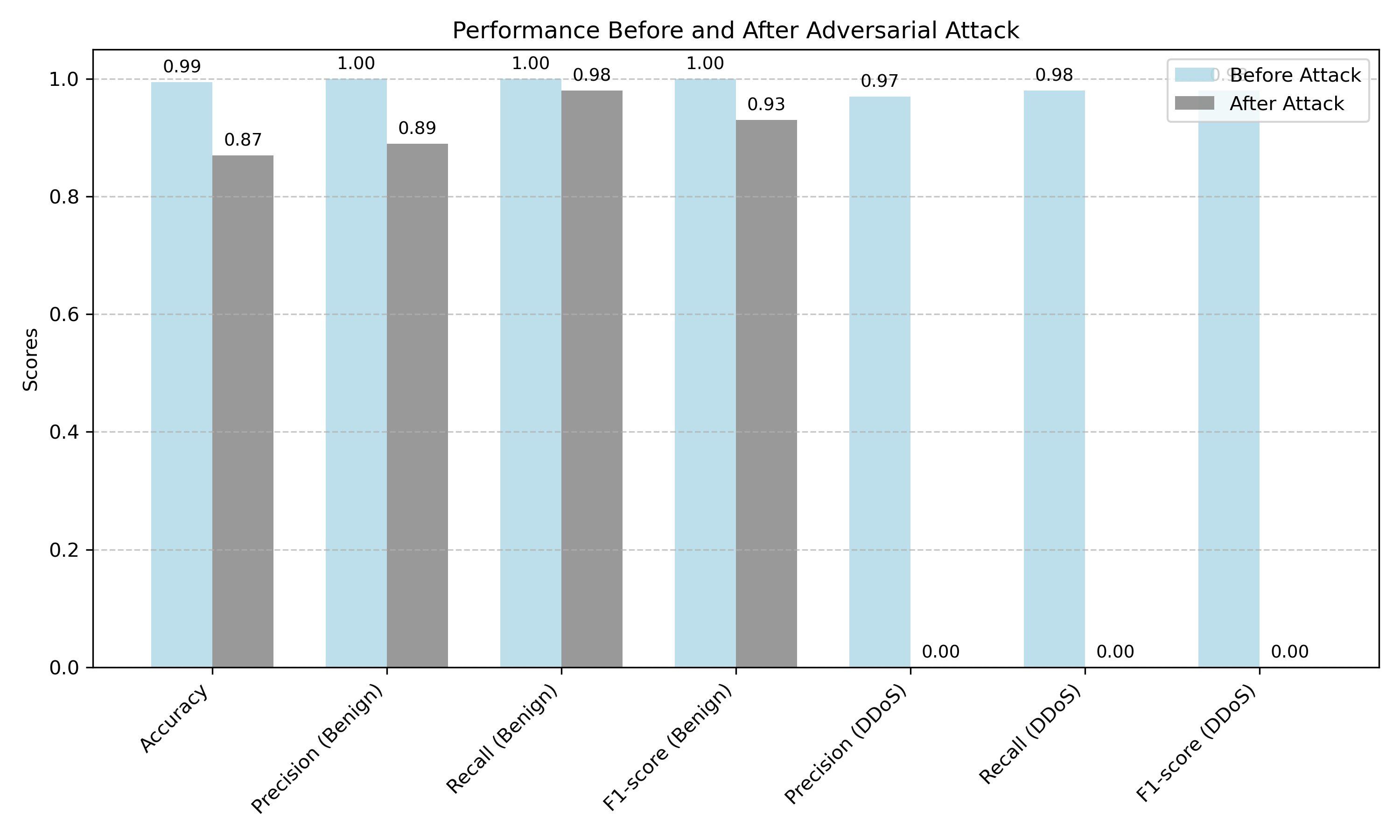} 
    \caption{Performance Analysis of the RF Model Before and After the Transfersability-based Attack}
    \label{fig:performance}
\end{figure*}

\section{Results and Discussion}
\label{sec:discussion}
Existing literature often ignores the alignment between academic research and real-world industry needs. To address this gap, we reproduced one of the leading studies in the field \cite{debicha2023adv}, critically examining its methodology and demonstrating how certain assumptions restrict practical applicability. Our purpose is not only to evaluate the feasibility of existing transferability-based attacks under real-world black-box settings but also to enhance the effectiveness of these approaches by raising awareness about the unrealistic assumptions they are built on. In order to achieve this, we propose a novel metric, named the Transferability Feasibility Score (TFS), that quantifies the level to which existing black-box attacks rely on unrealistic surrogate-target similarity; an assumption that is unlikely to hold in real-world scenarios.


Figure \ref{fig:performance} presents the performance metrics of the target RF-based NIDS before and after adversarial perturbations transferred by the surrogate model described in Section \ref{sec:methododlogy}. 
Before the adversarial attack, the RF model performs exceptionally well  with an accuracy of approximately \textbf{0.99}, demonstrating its resilience in identifying both benign and malicious traffic on a clean dataset. However, after the attack, the accuracy significantly drops to \textbf{0.87}, indicating the effectiveness of adversarial perturbations in reducing the model's reliability.




\textbf{Performance on benign traffic:} Even after the attack, the metrics for benign traffic—precision, recall, and F1-score—remain increased, with precision maintaining at \textbf{0.89}, recall at \textbf{0.98}, and F1-score at \textbf{0.93}. This stability implies that the attack focuses mainly on the detection of malicious traffic, with no impact on benign classifications.

\begin{figure*} [h!]
    \centering
    \includegraphics[width=0.98\textwidth]{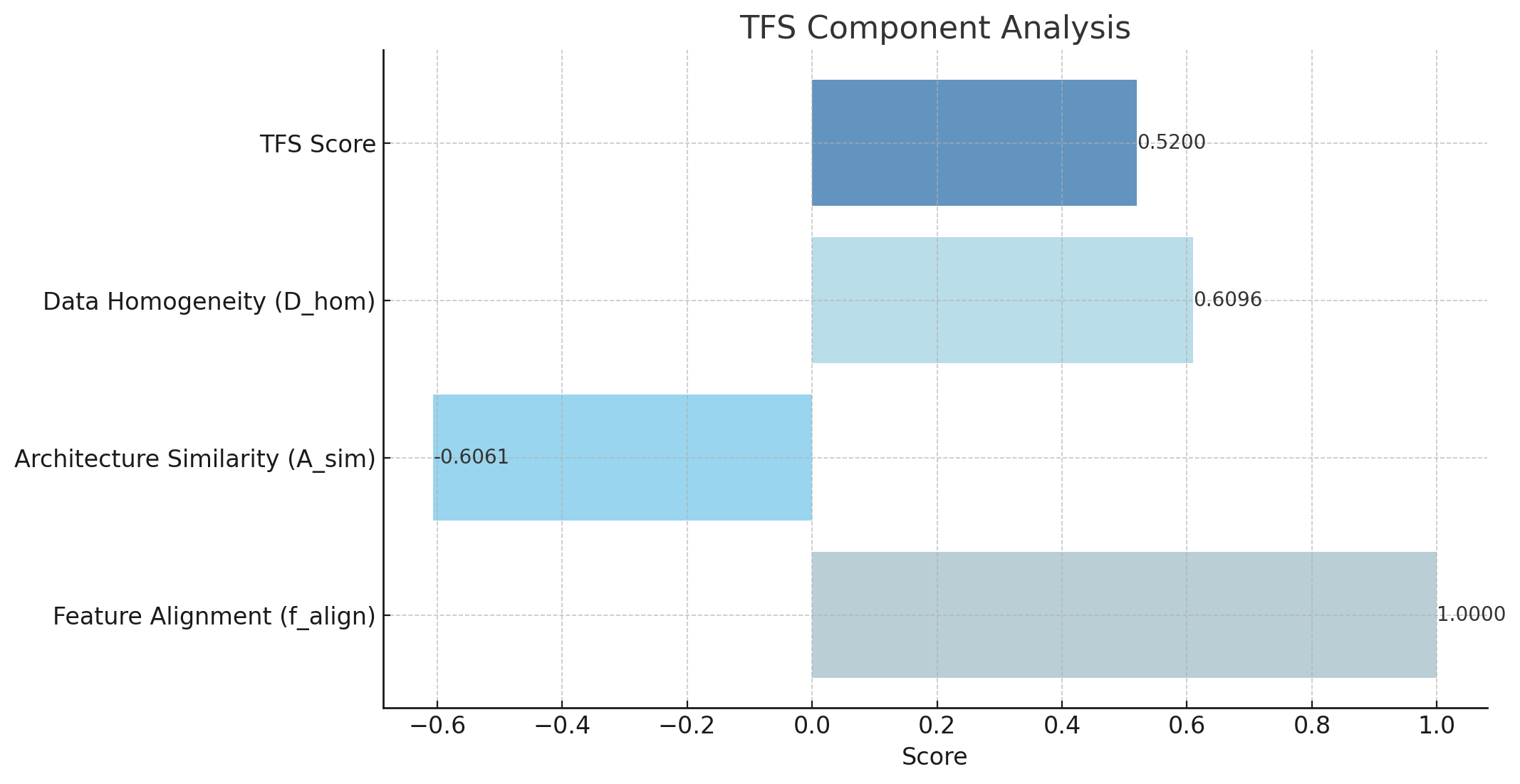} 
    \caption{Transferability Feasibility Score Analysis}
    \label{fig:TFS}
\end{figure*}

\textbf{Performance on DDoS attacks:} On the other hand, the DDoS traffic metrics completely collapse. Following the attack, the adversarial samples effectively tricked the model into totally failing to detect the intended class, as seen by the precision, recall, and F1-score for DDoS attacks all dropping to \textbf{0.00}. This illustrates how poorly the model can manage adversarial perturbations for malicious traffic.

While the findings demonstrate the success of the adversarial attack, they also raise questions about how feasible this effectiveness is under real-world conditions. Our proposed Transferability Feasibility Score (TFS) provides a structured solution for evaluating the practicality of transferability approach by quantifying how much the attack relies on similarity between the surrogate and target models in terms of input features, model architecture, and training data; an unrealistic condition in real-world black-box settings, where models could be dissimilar.


\textbf{Transferability analysis using TFS:} Figure \ref{fig:TFS} depicts a comprehensive analysis of the studied adversarial attack transferability through the TFS framework, evaluating the three crucial dimensions (i.e., feature alignment, architecture similarity and data homogeneity). This analysis presents a quantitative approach to evaluating transferability and overcoming the gap between theoretical attacks and practical applicability.
\begin{itemize}
\item \textbf{Feature alignment (\(f_{\text{align}}\)):}  The feature spaces of the surrogate and target models indicate high overlap, stemming from a strong score of \textbf{1.0}. This demonstrates that adversarial attacks can increase their potential for transferability by exploiting similarities in feature representation.
\item \textbf{Architecture similarity (\(A_{\text{sim}}\)):} Despite the reliance on similar feature spaces, the significant differences between the surrogate (MLP) and target (RF) models presents a considerable obstacle. The negative score of \textbf{-0.6061} indicates an architectural divergence, which means that the internal structural differences between the surrogate model and the target model reduce the potential for effective transferability. 
\begin{takeawaybox}
The negative value arises because the similarity measure is based on the Euclidean distance between the model parameters. By ensuring scale invariance through division by $\|P_t\|_2$, models with different parameter magnitudes can be fairly compared. A negative score presents a significant mismatch rather than similarity, highlighting that architectural differences present a serious challenge to transferability even when feature alignment and data homogeneity are high.    
\end{takeawaybox}

However, this architectural difference increases the attack's real-world feasibility because the surrogate and target models are unlikely to have the same architectures in practical scenarios.

\item \textbf{Data homogeneity (\(D_{\text{hom}}\)):} The moderate similarity score of \textbf{0.6096} indicates that the training data for the surrogate and target models share some characteristics. This component highlights the critical role of dataset alignment in supporting transferability.
\end{itemize}

It's important to note that the Wasserstein distance's sensitivity to outliers and high-dimensional feature spaces is a reflection of realistic variability rather than a limitation.   Transferability feasibility is by nature reduced by outliers and data complexity, which aligns with TFS's aim of measuring the structural compatibility of target and surrogate models. Consequently, this sensitivity reinforces the accuracy of TFS rather than decreasing its reliability.

\textbf{Overall TFS score:} The overall TFS score of \textbf{0.5200} proves a medium level of feasibility for adversarial attack transferability in this setup. Although it shows a moderate chance for practical application, additional development may increase its efficacy in a variety of difficult circumstances, increasing its overall viability in real-world situations.

\begin{remarkbox}
Since TFS was not created for real-time IDS deployment; but, to evaluate the practicality of transferability-based attacks under realistic structural constraints, its computational cost was not the focus of our study. TFS computation, which includes architectural similarity (Euclidean distance), feature alignment (Jaccard index), and data homogeneity (Wasserstein distance), was effective and did not introduce significant delays.
\end{remarkbox}

These findings demonstrate that even under optimal conditions in specific areas, such as feature alignment and data homogeneity, transferability is not guaranteed. A moderate TFS score, which indicates the possibility of transferability in situations with aligned feature spaces, is a result of strong feature similarity. However, the constraints created by architectural variations reinforce the necessity of evaluating transferability techniques in practice. By using the TFS framework, researchers can better understand the conditions and limitations that affect the transferability of adversarial attacks, which helps inform the development of more realistic and resilient adversarial strategies.

\begin{takeawaybox}
The reproduced paper \cite{debicha2023adv} did not consider an adversarial approach against ensemble learning methods, which with their multiple decision-making processes, are more resistant to transferability-based adversarial attacks \cite{pang2019improving}. 
\end{takeawaybox}

Based on the findings discussed above, particularly the limitations of existing adversarial strategies in considering industrial needs and the complexities of real-world scenarios, we propose the following recommendations to direct future research and development initiatives.

\begin{itemize}
\item \textbf{Reducing dependence on feature alignment}\\ 
Perfect feature alignment (\( f_{\text{align}} = 1 \)) is an idealized concept that is rarely valid in practical situations. When the feature sets of the target and surrogate models are different, adversarial strategies that mostly rely on feature overlap may not work. To address this:
\begin{itemize}
    \item Created adversarial behaviors have to adapt to systems with slight or partial feature overlap.
    \item High-level representations must be exploited or transferable patterns that are less dependent on specific features.
\end{itemize}

\item \textbf{Enhancing adversarial transferability across diverse architectures} \\
Low architecture similarity (\( A_{\text{sim}} \)) presents a serious challenge to transferability, as structural differences between surrogate and target models can limit the effectiveness of adversarial examples. To enhance robustness:
\begin{itemize}
    \item Reliance on ensemble learning-based models as surrogates and target models to approximate different architectural patterns.
    \item Examination of attack techniques like decision-based or gradient-free attacks that are independent of particular architectures.
\end{itemize}

\item \textbf{Adaptation to data distribution variations} \\
Dissimilarities in data distributions (\( D_{\text{hom}} \)) between surrogate and target models minimize the generalizability of adversarial examples. To cope with this, it is recommended to use data augmentation or domain adaptation strategies to make adversarial examples work in a variety of distributions.

\item \textbf{Evaluation of practical applicability} \\
The proposed approaches need to be evaluated in realistic black-box scenarios, such as limited query access, restricted knowledge of the target model (e.g., using side-channel indicators), reliance on realistic datasets, and feasibility metrics like the proposed TFS for transferability-based attacks.
\end{itemize}
These insights show the importance of refining adversarial attack strategies that are not only effective but also adaptable to real-world complexities, such as diverse network architectures and domain-specific data. Such steps will contribute to the development of robust and practical solutions for real-world cybersecurity challenges.

\section{Conclusion}\label{sec:conclusions}
This paper proposes a novel benchmark for evaluating the transferability of adversarial attacks, namely Transferability Feasibility Score (TFS). 
TFS evaluates three key factors: feature alignment, which quantifies the overlap between the target and surrogate model feature spaces; architecture similarity between models; and data homogeneity, which assesses how comparable training datasets are. These are necessary for understanding the feasibility of transferring adversarial attacks.
The findings show that transferability is not always feasible, particularly against robust models such as ensemble learning methods. This study serves as a guide for future research, encouraging the development of more practical adversarial strategies and adaptive defenses to address real-world challenges in intrusion detection systems.

\bibliographystyle{unsrt}

\end{document}